\documentclass[prl,twocolumn,showpacs]{revtex4}
\usepackage{epsfig,amsmath}

\begin{document}

\title{Study of two-dimensional electron systems in the renormalized-ring-diagram approximation} 

\author{Xin-Zhong Yan$^{1,2}$, and C. S. Ting$^1$}
\affiliation{$^{1}$Texas Center for Superconductivity, University of 
Houston, Houston, TX 77204\\
$^{2}$Institute of Physics, Chinese Academy of Sciences, P.O. Box 603, 
Beijing 100080, China}

\date{\today}
 
\begin{abstract}

With a super-high-efficient numerical algorithm, we are able to self-consistently calculate the Green's function in the renormalized-ring-diagram approximation for a two-dimensional electron system with long-range Coulomb interactions. The obtained ground-state energy is found to be in excellent agreement with that of the Monte Carlo simulation. The numerical results of the self-energy, the effective mass, the distribution function, and the renormalization factor of the Green's function for the coupling constants in the range $0 \le r_s \le 30$ are also presented.
\end{abstract}


\maketitle

Two-dimensional electron systems (2DES) with long-range Coulomb interactions, realized in the semiconductor heterostructures and inversion layers \cite{Ando}, continuously attract a lot of attentions despite it has been studied for more than three decades. Most of the theoretical calculations so far were based on the random-phase approximation (RPA) \cite{Ting,Rajagopal,Hu,Sarma}, which is expected to be accurate in the weak coupling or high density regime. For taking into account the strong-coupling effect beyond the PRA, the usual way is to adopt the local-field correction (LFC) to the RPA \cite{Singwi,Utsumi,Iwamoto,Jang,Yarlagadda,Chen}. In all these works, the Green's functions appeared in the diagrams of the standard RPA as well as the one with the LFC are not renormalized. A crucial condition for a better approximation is that the renormalized Green's function should satisfy some microscopic conservation laws, otherwise the approximation may lead to unphysical consequences \cite{Baym}. Among various approximations, the renormalized-ring-diagram approximation (RRDA), in which the Green's function needs to be self-consistently determined from the relevant integral equations, is well known to satisfy this condition \cite{Baym} and thus should be a sound approach. Because tremendous numerical efforts were needed to solve the integral equations in RRDA, the physics in this approach had only been studied for small coupling constants \cite{Faleev}, and there existed no solutions to RRDA from the intermediate to strong-coupling regimes. To develop a numerical scheme to solve RRDA equations for the purpose of understanding 2DES in a wider range of couplings is still a challenging problem in modern many-particle physics. In this paper, we use a super-high-efficient algorithm to solve the RRDA equations. With the solutions, the ground-state energy is obtained, and it is in excellent agreement with the result of the fixed-node-diffusion Monte Carlo (MC) simulation \cite{Attac}. In addition, the self-energy of the Green's function, the effective mass, the single particle distribution function as well as the renormalization factor of the Green's function are also calculated. 

We consider a 2DES with density of $n$ at temperature $T$ embedded in an uniform neutralizing background of positive charge. Through out this paper, we will use the units in which $\hbar = k_B = m = a = 1$ (with $m$ the mass of the electron, and $a$ the Wigner-Seitz radius of an electron in the 2DES). The system is characterized by two dimensionless parameters: $\theta = T/E_F$ and $r_s = a/a_B$ where $E_F=\pi n$ (= 1 in our units) is the Fermi energy and $a_B$ the Bohr's radius. The electronic Green's function $G$ is related to the self-energy $\Sigma$ via
\begin{eqnarray}
G(\vec k,i\omega_n)= [i\omega_n-\xi_{\vec k}-\Sigma(\vec 
k,i\omega_n)]^{-1}, \label{G}
\end{eqnarray}
where $\xi_k = \epsilon_k-\mu$ with $\epsilon_k = k^2/2$ and $\mu$ the chemical potential, and $\omega_n$ is the fermionic Matsubara frequency. For brevity, we hereafter will use $k \equiv (\vec k, i\omega_n)$ for the argument unless stated otherwise. By the RRDA \cite{Baym}, the self-energy $\Sigma$ is given by
\begin{equation}
\Sigma(k)=-\frac{1}{V\beta}\sum_{q}G(k+q)\frac{V(q)}{1-V(q)\chi(q)}, \label{se}
\end{equation}
where $\beta = 1/T$, $V(q)=2\pi r_s/q$ is the Coulomb interaction, $V$ is the area of the system, and
\begin{equation}
\chi(q) = \frac{2}{V\beta}\sum_{k}G(k)G(k+q)  
\label{chi} 
\end{equation}
is the electron polarizability, $q \equiv (\vec q, i\Omega_m)$ with $\Omega_m$ the bosonic Matsubara frequency. The chemical potential $\mu$ is determined by the electron density,
\begin{equation}
n = \frac{2}{V\beta}\sum_{k}G(k)\exp(i\omega_n0^+). \label{mu} 
\end{equation}
By solving Eqs. (\ref{G})-(\ref{mu}), the functions $G$ and $\Sigma$, and the chemical potential $\mu$ can be self-consistently obtained. 

In the numerical procedure, the most time-consuming computations are calculations of $\Sigma$ and $\chi$ because of the summation over the Matsubara frequency and the integration over the momentum. We have developed an effective algorithm for doing such a computation. In this method, the summation is taken only over some selected Matsubara frequencies distributed in $L$ successively connected blocks each of them containing $M$ frequencies, with each term under the summation multiplied by a weighting factor. For details of this algorithm, the reader is referred to Ref. \onlinecite{Yan}. In the present calculation, we have used the parameters $[h,L,M] = [2,15,5]$ for selection of the Matsubara frequencies, where $h$ is the integer-parameter that the stride in the $\ell$-th block is $h^{(\ell-1)}$. The total number of the selected frequencies is $L(M-1)+1 = 61$ with the largest number $N \sim 2^L(M-1) = 2^{17}$ for the cutoff frequencies $\Omega_N=2N\pi T$ and $\omega_N = (2N-1)\pi T$. We will see later that the 61 Matsubara frequencies are sufficient to describe the self-energy and thereby the Green's function. Instead of summing over $N$ Matsubara frequencies by a usual method, we here take the summation only over the 61 ones. Therefore, the efficiency of the algorithm for the present calculation is $N/61 \sim 2^{11}$! For the lowest temperature considered here, $\theta = 0.03$, we have $\omega_N/E_F \sim 2.47\times 10^4$. In our calculation, the largest $\epsilon_k$ is $\epsilon_M = k^2_M/2$ with $k_M = 50$. Therefore, the cutoff $\omega_N/\epsilon_M \sim 20$ is sufficiently large. 

\begin{figure} 
\centerline{\epsfig{file=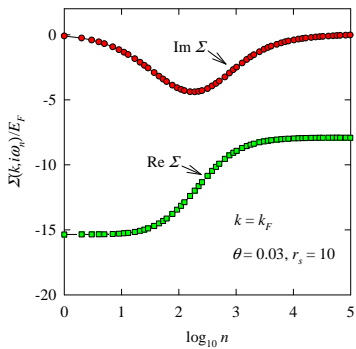,width=5.5 cm}}
\caption{(Color online) Self-energy $\Sigma(k,i\omega_n)$ as function of $n$ at $k = k_F$, $\theta = 0.03$, and $r_s = 10$.}\label{selfe}
\end{figure}

On the other hand, the momentum-space convolution integrals in Eqs. (\ref{se}) and (\ref{chi}) can be easily performed by Fourier transforming into real space. For illustrating the two-dimensional Fourier transform, we here discuss the transformation of the Green's function. In the real space, the Green's function is given by 
\begin{equation}
G(r,i\omega_n) = \int_0^{\infty}\frac{dk}{2\pi}kG(\vec k,i\omega_n)J_0(kr) \label{Gr} 
\end{equation}
where $J_0$ is the first kind Bessel function of order zero. However, Eq. (\ref{Gr}) is not in a favorite form for the numerical computation since $kG(\vec k,i\omega_n)$ behaves like $O(1/k)$ at $k \to \infty$ and $j_0(kr)$ is an oscillatory function; the integrand decreases so slowly at large $k$ that a precise numerical result is hardly obtained. By observing the asymptotic behavior, 
\begin{equation}
G(k,i\omega_n) \to G^0(k,i\omega_n), ~~~~{\rm at}~ k \to \infty~{\rm or}~n \to \infty \label{as} 
\end{equation}
with $G^0(\vec k,i\omega_n)=1/(i\omega_n+\mu_0-k^2/2)$ and $\mu_0$ respectively the Green's function and the chemical potential of the free electrons, we choose $G^0$ as the auxiliary function for $G$. The Fourier transform of $G^0(\vec k,i\omega_n)$ is given by
\begin{equation}
G^0(r,i\omega_n) = -K_0(pr)/\pi \label{Gr0} 
\end{equation}
where $K_0$ is the second kind modified Bessel function of order zero, and $p = (-2\mu_0-i2\omega_n)^{1/2}$. Thus, equation (\ref{Gr}) can be rewritten as
\begin{equation}
G(r,i\omega_n) = \int_0^{\infty}\frac{dk}{2\pi}k\delta G(\vec k,i\omega_n)J_0(kr) -K_0(pr)/\pi       \label{Grr} 
\end{equation}
with $\delta G(\vec k,i\omega_n)=G(\vec k,i\omega_n)-G^0(\vec k,i\omega_n)$. Now, the integrand in Eq. (\ref{Grr}) drops fastly with $k\delta G(\vec k,i\omega_n)\sim O(k^{-3})$ at $k \to \infty$. Equation (\ref{Grr}) can be further reformed for dealing with the fast oscillatory behavior of $J_0(kr)$ at large $r$. For doing this, we consider the integral
\begin{equation}
g(r) = \int_0^{\infty}dk f(k)J_0(kr).  \label{A1} 
\end{equation}
For $J_0(kr)$, we choose the auxiliary function as
\begin{eqnarray}
A(kr) &=& \sqrt{\frac{c}{kr+c}}\{\sin(kr)+\cos(kr)\nonumber \\
& & +\frac{kr}{8(kr+c)^2}[\sin(kr)-\cos(kr)]\}  
\end{eqnarray}
with $c=1/\pi$. $A(kr)$ has the same asymptotic behavior as $J_0(kr)$ at $kr \to \infty$. In addition, $A(0) = J_0(0) = 1$. Equation (\ref{A1}) can be rewritten as
\begin{eqnarray}
g(r) &=& \int_0^{\infty}dk f(k)[J_0(kr)-A(kr)]   \nonumber\\ 
& & +\sqrt{c}\int_0^{\infty}dk\frac{f(k)}{\sqrt{kr+c}}[1-\frac{kr}{8(kr+c)^2}]\cos(kr)\nonumber\\
& & +\sqrt{c}\int_0^{\infty}dk\frac{f(k)}{\sqrt{kr+c}}[1+\frac{kr}{8(kr+c)^2}]\sin(kr) .\nonumber
\end{eqnarray}
Now, the first integral in the above equation can be integrated by the simple numerical method since the fast oscillatory behavior of $J_0$ is considerably cancelled by $A(kr)$. On the other hand, the second and third Fourier integrals can be numerically integrated using the Filon's method. In our calculation, the momentum integration is over the range $0 \le k \le 50$. Because the electron distribution function varies drastically around the Fermi momentum $k_F = \sqrt{2}$, we divide the $k$-range into three segments: $[0, k_F-k_0]$, $[k_F-k_0,k_F+k_0]$, and $[k_F+k_0,50]$ with $k_0 = \sqrt{8\theta}/(1+\sqrt{32\theta})$. The $k$-integrals are calculated with 200 meshes in each segment. In the real space, $r$ is confined to $0 \le r \le 80$, which is divided into three segments: $[0, 5]$, $[5,30]$, and $[30,80]$, with 300 meshes in each segment.

In the above example, the accuracy and efficiency of numerical integration is improved by choosing the proper auxiliary function. Analogously, we can choose the auxiliary function for the summation over the Matsubara frequency as well. For example, the auxiliary function for calculating $\chi(q)$ is chosen as $G^0(k)G^0(k+q)$ that is the asymptotic limit of $G(k)G(k+q)$ at $\omega_n \to \infty$ or $k \to \infty$. On the other hand, the free-particle polarizability $\chi^0(q)$ can be easily obtained by
\begin{equation}
\chi^0(\vec q,i\Omega_m) = \frac{\beta}{4\pi}\int_0^{\infty}d\epsilon\frac{Y(q,\Omega_m)-1}{\cosh^2[(\epsilon-\mu_0)/2T]}      \label{chi0} 
\end{equation}
with $Y(q,\Omega_m)=(\sqrt{x^2+y^2}+x)^{1/2}$ and $x=1/2-\Omega^2_m/2\epsilon_q^2-2\epsilon/\epsilon_q$, $y= \Omega_m/\epsilon_q$. 

By iteration, we have solved Eqs. (\ref{G})-(\ref{mu}) for $0 < r_s \le 30$ at finite temperatures. Firstly, in order to illustrate the validity of our selection for the Matsubara frequencies, we show in Fig. \ref{selfe} the result for the self-energy $\Sigma(k,i\omega_n)$ for $k = k_F$, $\theta = 0.03$, and $r_s = 10$. The symbols represent the values of $\Sigma(k,i\omega_n)$ calculated at the selected $n$'s. This is a typical result of $\Sigma(k,i\omega_n)$ in the strong-coupling regime. As seen from Fig. \ref{selfe}, the selected frequencies are sufficient for describing $\Sigma(k,i\omega_n)$; though the selected $\omega_n$'s are sparsely distributed at large $n$, meanwhile, the function $\Sigma(k,i\omega_n)$ varies slowly. At $\omega_n \to \infty$, the imaginary part of $\Sigma(k,i\omega_n)$ vanishes, but the real part remains finite. Actually, at $\omega_n \to \infty$, we get $\Sigma(k,i\omega_n) \to $ the exchange part of the self-energy. 

\begin{figure} 
\centerline{\epsfig{file=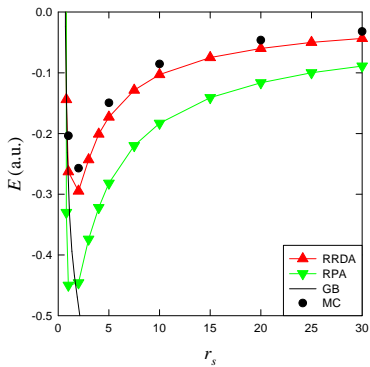, width=5.5 cm}}
\caption{(Color online) Ground-state energy in atomic unit (a.u.) as function of $r_s$. The present result (RRDA) is compared with that of the RPA, GB \cite{Rajagopal}, and MC \cite{Attac}.}\label{gse}
\end{figure}

In Fig. \ref{gse}, the ground-state energy per electron in atomic unit is shown as a function of $r_s$. The results of present calculation are obtained by extrapolation of the finite-temperature values. Our main result denoted by RRDA is compared with the RPA, the high-density expansion \cite{Rajagopal} by the scheme of Gell-Mann and Brueckner (GB) \cite{GB}, and the MC simulation \cite{Attac}. In the weak coupling regime, both the RRDA and RPA reproduce the GB result very well. However, for strong coupling, the RRDA is much closer to the MC than the RPA. 

\begin{figure} 
\centerline{\epsfig{file=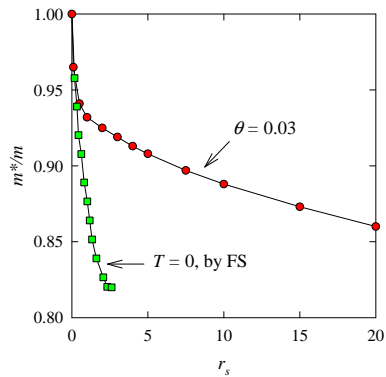, width=5.5 cm}}
\caption{(Color online) Effective mass $m^*/m$ as function of $r_s$. The present result at $\theta = 0.03$ is compared with that at $\theta = 0$ obtained in Ref. \onlinecite{Faleev} (FS) with a different numerical method.}\label{ems}
\end{figure}

Shown in Fig. \ref{ems} is the effective mass $m^*/m$ as a function of $r_s$ at $\theta = 0.03$. The values calculated by Faleev and Stockman (FS) with a different numerical scheme for $T = 0$ is also depicted for comparison. The present result shows that $m^*/m$ is a monotonically decreasing function of $r_s$ in a wide range of $r_s$. This is different from the RPA (not shown here) by which $m^*/m$ decreases at small $r_s \le 0.1$, but then increases with $r_s$ \cite{Sarma}. The present result on $m^*/m$ seems to be contradictory to the explanation to the experiments on the temperature dependence of the Shubnikov-de Haas oscillations in a 2DES \cite{Shashkin}. The experiments were explained as $m^*/m$ increasing with $r_s$. To this point, there may be two possibilities: (1) the RRDA may not be good enough, or (2) the experimental results may require a different interpretation; for example, the conductivity formula to fit the experiments was based on non-interacting electrons except the mass is assumed to be renormalized. Further investigation is needed to get a clear understanding of this problem. 
 
\begin{figure} 
\centerline{\epsfig{file=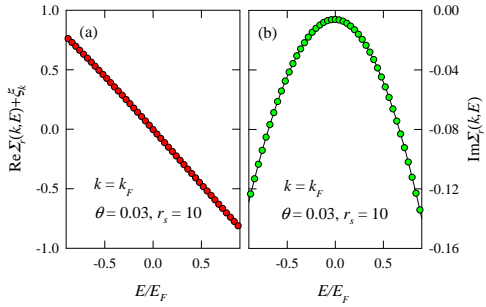, width=7. cm}}
\caption{(Color online) Self-energy $\Sigma_r(k,E)$ as function of $E$ at $k = k_F$ for $\theta = 0.03$ and $r_s =10$. (a) ${\rm Re}\Sigma_r(k_F,E)+\xi_{k_F}$ and (b) the imaginary part.}\label{slfre}
\end{figure}

The effective mass $m^*/m$ has been extensively investigated by many calculations using the RPA. With the RPA self-energy, $m^*/m$ is calculated by the on-shell scheme \cite{Ting}, which is not a self-consistent approximation for the quasiparticle energy. Contrary to the RPA, the RRDA is a conserving approximation for the single-particle Green's function \cite{Baym}; the Green's function satisfies the Luttinger theorem \cite{Luttinger1,Luttinger} by which the Fermi surface is unchanged for the 2DES. To see this, we show in Fig. \ref{slfre} the retarded self-energy $\Sigma_r(k,E)$ as function of $E$ at $k = k_F$ for $\theta = 0.03$ and $r_s =10$. $\Sigma_r(k,E)$ is obtained by the Pad\'e approximation \cite{Vidberg}. From Fig. \ref{slfre}(a), it is seen that ${\rm Re}\Sigma_r(k_F,E)$ is essentially a linear function of $E$ in the neighborhood of $E = 0$. It is clear that ${\rm Re}\Sigma_r(k_F,0)+\xi_{k_F} = 0$, which means that the quasiparticle energy $E$ vanishes at the Fermi surface consistent with the Luttinger theorem. The equality ${\rm Re}\Sigma_r(k_F,0)+\xi_{k_F} = 0$ originates from the fact that ${\rm Re}\Sigma_r(k_F,0)$ is compensated by the shift of the renormalized $\mu$ from $\mu_0$. The imaginary part shown in Fig. \ref{slfre}(b) has a parabola form with a small value at $E = 0$. At $T = 0$, ${\rm Im}\Sigma_r(k_F,E) \propto -E^2$ at small $E$ \cite{Luttinger1}. The small negative value of ${\rm Im}\Sigma_r(k_F,0)$ shown in Fig. \ref{slfre}(b) is a finite-temperature effect. (All these facts indicate that our numerical computation is very accurately performed.) On the other hand, since the bare chemical potential $\mu_0$ is used in the RPA calculation, ${\rm Re}\Sigma_r(k_F,0)$ cannot be compensated and thereby the quasiparticle energy does not vanishes at $k_F$.

\begin{figure} 
\centerline{\epsfig{file=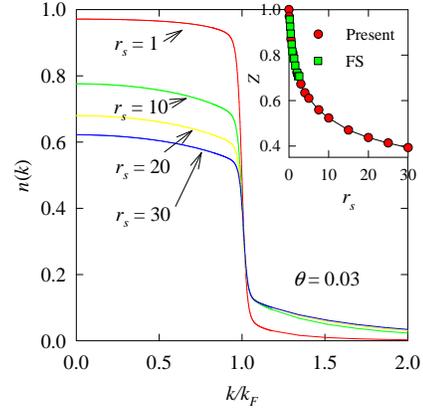, width=6. cm}}
\caption{(Color online) Distribution function $n(k)$ at $\theta = 0.03$ and various $r_s$.
The inset is the $Z$-factor as function of $r_s$. The circles represent the results of the present calculation, and the squares are that of Ref. \onlinecite{Faleev}.}\label{zf}
\end{figure}

In Fig. \ref{zf}, we exhibit the result for the distribution function $n(k)$ at $\theta = 0.03$ for various $r_s$. It is well known that $n(k)$ at $r_s = 0$ is the Fermi-Dirac distribution function $F^0(k)$ for the free electrons. At very small $r_s$, $n(k)$ is close to $F^0(k)$. With increasing $r_s$, $n(k)$ becomes very different from $F^0(k)$; it is gradually suppressed at $k < k_F$ while growing up at $k > k_F$. At the Fermi momentum, $n(k)$ decreases dramatically, but seems continuously. At zero temperature, $n(k)$ should have a discontinuity at $k = k_F$. The continuous behavior appeared in Fig. \ref{zf} is attributed to the finite-temperature effect by which the discontinuity is rounded. Shown in the inset is the renormalization factor $Z$ of the Green's function. $Z$ is associated with the abrupt drop in $n(k)$ of the ground state at $k_F$. Within the range of $r_s$ studied here, the magnitude of $Z$ is consistent with the sharp drop in $n(k)$ at $k_F$. For comparison, the result by FS is also shown in the inset. By their method \cite{Faleev}, they could solve the RRDA equations only within $r_s \le 2.62$. Clearly, their result is reproduced by the present calculation.

In summary, we have solved the RRDA equations for 2DES at $r_s \le 30$ with our super-high-efficient numerical algorithm. The obtained ground-state energy is in excellent agreement with the Monte Carlo result. For the effective mass $m^*/m$, the RRDA result shows $m^*/m$ is a decreasing function of $r_s$, which is considerably different from the RPA. The distribution function $n(k)$ and the renormalization factor $Z$ of the Green's function are calculated for a wide range of $r_s$.

The authors thank Dr. Q. Wang for useful discussion on the related problems. This work was supported by a grant from the Robert A. Welch Foundation under No. E-1146, the TCSUH, and the National Basic Research 973 Program of China under grant number 2005CB623602.

\end{document}